# Local Term Weight Models from Power Transformations

## Development of BM25IR: A Best Match Model based on Inverse Regression

*Abstract*–In this article we show how power transformations can be used as a common framework for the derivation of local term weights. We found that under some parametric conditions, BM25 and inverse regression produce equivalent results. As a special case of inverse regression, we show that the largest increment in term weight occurs when a term is mentioned for the second time. A model based on inverse regression (BM25IR) is presented. Simulations suggest that BM25IR works fairly well for different BM25 parametric conditions and document lengths.

Keywords: box-cox, tukey, bm25, bm25ir, inverse regression, local weights, term weights, document lengths

© E. Garcia; admin@minerazzi.com

## Introduction

The weight of a term $i$ in a document $j$ with an occurrence frequency $f_{i,j}$ is computed by assigning a local ($L_{i,j}$), global ($G_i$), and normalization ($N_j$) weight. Several models for doing this have been proposed, some using a combination of assumptions, intuitions and experimental observations (Chisholm &Kolda, 1999; Jones & Furnas, 1987; Lee, Chuang, & Seamons, 1997; MacFarlane, 2001; Robertson, 2004; Robertson & Walker, 1994; Robertson, Walker, Jones, Hancock-Beaulieu, & Gatford, 1994; Robertson & Zaragoza, 2009; Salton & Buckley, 1987; Salton, Wong, & Yang, 1975; Salton & Yang, 1973; Sanderson & Ruthven, 1996).

However no common framework is given for their systematic derivations. The purpose of this article is to present such a framework based on power transformations. We show that many of the local weights models found in the literature, and new ones, can be derived in this way.

## Why Power Transformations

Typically there are three reasons for modifying a data set through power transformations:

- To make the distribution of a data set closer to that of a normal distribution.
- To linearize the relationships between variables.
- To stabilize the variance.

While it is true that word occurrences in documents have been modeled as belonging to Poisson mixtures, it is also true that good keywords are far from Poisson (Church & Gale, 1995a; 1995b).

Although power transformations ensure that the assumption for linearity, normality, and homoscedasticity hold, the main objective is to make inferences on the power transformation parameter, even in cases where no power-transformation brings a distribution to normality (Li, 2005). It is in this context that power transformation methods are used in the present article.

## Tukey and Box-Cox Power Transformations

The best known power transformation models are due to Tukey (1957) and Box & Cox (1964).

Tukey:

$$y^* = (y + \lambda_2)^{\lambda_1} \tag{1}$$

Box-Cox:

$$y^* = \frac{(y+\lambda_2)^{\lambda_1} - 1}{\lambda_1} \tag{2}$$

where

*y* is a numerical value

*y** is a transformed value,

$\lambda_1$is a power parameter that can adopt any real value

$\lambda_2$ is a positive constant typically used to offset any negative or zero *y* value.

These transformations are very effective when the data do not describe an inflection point (Hossain, 2011; Steiger, 2009; Sakia, 1992). A comparative of these models is given in Table 1.

**Table 1. Common data transformation models.**

| Model | Condition | $\lambda_1 \neq 0$ | $\lambda_1 = 0$ |
|---|---|---|---|
| **Tukey** | $y + \lambda_2 > 0$ | $y^* = (y + \lambda_2)^{\lambda_1}$ | $y^* = \ln(y + k)$ |
| | $y > 0,\ \lambda_2 = 0$ | $y^* = y^{\lambda_1}$ | $y^* = \ln(y)$ |
| **Box-Cox** | $y + \lambda_2 > 0$ | $y^* = \frac{(y + \lambda_2)^{\lambda_1} - 1}{\lambda_1}$ | $y^* = \ln(y + k)$ |
| | $y > 0,\ \lambda_2 = 0$ | $y^* = \frac{y^{\lambda_1} - 1}{\lambda_1}$ | $y^* = \ln(y)$ |

The difference between the models is that for $\lambda_1 \neq 0$ Box-Cox's model shifts by -1 and normalize with $\lambda_1$ the scales. For $\lambda_1 = 1$ and $\lambda_2 = 0$, Tukey's model does not change the data, but Box-Cox's subtracts 1. This does not change the results, though.

For $\lambda_1 = 0$, both models return logs, but by different means: In the Tukey model, the derivative $dy^*/d\lambda_1$ at $\lambda_1 = 0$ is evaluated where in the Box-Cox model, the *l'Hôpital's Rule* is applied. In both cases the base of the logarithms does not matter.

Among others, the following transformations are obtained from both models by setting $\lambda_1$:

- quadratic ($\lambda_1 = 2$)
- linear ($\lambda_1 = 1$)
- square root ($\lambda_1 = 0.5$)
- logarithmic ($\lambda_1 = 0$)
- inverse or reciprocal ($\lambda_1 = -1$)
- inverse square root ($\lambda_1 = -0.5$)
- inverse quadratic ($\lambda_1 = -2$)

## Box-Cox Transformations in Information Retrieval

Nowadays Box-Cox transformations are preferred over Tukey's. So it is not surprising to see these applied in IR. For instance, Gerani, Zhai, & Crestani (2012) used the transformations in relevance ranking work. Molina, Torres-Moreno, SanJuan, Sierra, & Rojas-Mora (2013) applied Box-Cox transformations to low term frequencies. Lv & Zhai (2011) and Zhou (2014) have use these to overcome problems associated to document lengths in BM25 models.

However, at the time of writing, these transformations have not been used as a framework for systematically deriving local weights. This is precisely the purpose of the present article.

## Derivation of Local Weight Models

Some of the models derived below are discussed by Chisholm & Kolda (1999) and in a previous tutorial (Garcia, 2016). For consistency sake with those reports, we adopt the following conventions. $y^*$ is replaced with $L_{i,j}$, $y$ with $f_{i,j}$, $\lambda_1$ with $p$, $\lambda_2$ with $k$, and *ln* with *log* where *log* are base 2 logarithms, although the base used does not really matter. Table 2 lists the models derived from Tukey and Box-Cox power transformations for $p$ = 2, 1, ½, 0, -1/2, -1, and -2 where $k > 0$.

**Table 2. Local Term Weight Models Derived with (1) and (2)**

| *p* | Tukey Transformations, (1) | Box-Cox Transformations, (2) |
|---|---|---|
| **2** | $L_{i,j} = \left(f_{i,j} + k\right)^2$ | $L_{i,j} = 0.5\left(f_{i,j} + k\right)^2 - 0.5$ |
| **1** | $L_{i,j} = f_{i,j} + k$ | $L_{i,j} = f_{i,j} + k - 1$ |
| **1/2** | $L_{i,j} = \left(f_{i,j} + k\right)^{1/2}$ | $L_{i,j} = 2\left(f_{i,j} + k\right)^{1/2} - 2$ |
| **0** | $L_{i,j} = log\left(f_{i,j} + k\right)$ | $L_{i,j} = log\left(f_{i,j} + k\right)$ |
| **-1/2** | $L_{i,j} = \frac{1}{\left(f_{i,j} + k\right)^{1/2}}$ | $L_{i,j} = 2 - 2\frac{1}{\left(f_{i,j} + k\right)^{1/2}}$ |
| **-1** | $L_{i,j} = \frac{1}{f_{i,j} + k}$ | $L_{i,j} = 1 - \frac{1}{f_{i,j} + k}$ |
| **-2** | $L_{i,j} = \frac{1}{\left(f_{i,j} + k\right)^2}$ | $L_{i,j} = 0.5 - 0.5\frac{1}{\left(f_{i,j} + k\right)^2}$ |

## Derivation of Local Weight Models

A close look at Table 2 reveals that for $f_{i,j} + k = 1$ and $p \neq 0$, Tukey's model returns $L_{i,j} = 1$ whereas for $f_{i,j} + k = 1$ and any value of $p$, Box-Cox's model returns $L_{i,j} = 0$. Other combinations of $p$ and $k$ produce more interesting solutions.

For instance, the following term weight models (Chisholm & Kolda, 1999; Garcia, 2016) are derived from Table 2 after some slight modifications.

- **FREQ Model,** $L_{i,j} = f_{i,j}$. Derivable with
  - Tukey's, by setting $p = 1$ and $k = 0$.
  - Box-Cox's , by setting $p = k = 1$.
- **SQRT Model**, $L_{i,j} = 1 + \left(f_{i,j} - 0.5\right)^{1/2}$. Derivable with
  - Tukey's, by setting $p = ½$, $k = -½$, and adding 1.
- **LOGA Model,** $L_{i,j} = 1 + \ log\left(f_{i,j}\right)$. Derivable with
  - Tukey's and Box-Cox's, by setting $p = k = 0$ and adding 1.
- **LOGN Model**, $L_{i,j} = \frac{1+\log\left(f_{i,j}\right)}{1+\log\left(ave\,f_{i,j}\right)}$. Derivable with
  - Tukey's and Box-Cox's, by setting $p = k = 0$, adding 1, and normalizing the scale by dividing by $1 + \log\left(ave f_{i,j}\right)$
- **LOGLN Model**, $L_{i,j} = \frac{\log\left(f_{i,j}+1\right)}{\log\left(document\ length_j\right)}$ . Derivable with
  - Tukey's and Box-Cox's by setting $p = 0$, $k = 1$, and normalizing the scale by dividing by $\log\left(document\ length_j\right)$.
- **LOGG Model**, $L_{i,j} = 0.2 + 0.8 * \log\left(f_{i,j} + 1\right)$. Derivable with
  - Tukey's and Box-Cox's, by setting $p = 0$, $k = 1$, multiplying by 0.8 and adding 0.2

In the next sections we show that local weight components of Best Match (BM) algorithms can be derived from Box-Cox transformations. For some specific parametric conditions, Best Match (BM) algorithms and inverse regression produce equivalent results.

## BM Term Weight Components

The local term weight component of BM25 is defined as

$$L_{i,j} = \frac{f_{i,j}}{f_{i,j}+K} * (k_1 + 1) = \frac{f_{i,j}}{f_{i,j}+k_1\left((1-b)+b*\frac{dl_j}{avedl}\right)} * (k_1 + 1) \quad (3)$$

where

$dl_j$ = document length of document *j*

*avedl* = average document length

$K = k_1 B$ = attenuation factor

$B = (1-b) + b * \frac{dl_j}{avedl}$ = document length normalization function

*b* = normalization parameter

and where $(k_1 + 1)$ is a scaling factor.

For *b* = 1 which is the case of full document length normalization, a BM25 local weight reduces to a BM11 weight,

$$L_{i,j} = \frac{f_{i,j}}{f_{i,j}+k_1\left(\frac{dl_j}{avedl}\right)} * (k_1 + 1) \quad (4)$$

and to a BM15 weight for *b* = 0, corresponding to zero document length normalization,

$$L_{i,j} = \frac{f_{i,j}}{f_{i,j}+k_1} * (k_1 + 1) \quad (5)$$

Without loss of generality, for documents of average length $\frac{dl_j}{avedl} = 1$, $K = k_{1,}$ and the local weight components of BM11, BM15, and BM25 are the same.

## BM as an Inverse Regression Solution

In general, the relationship between $f_{i,j}$ and $L_{i,j}$ can be traced back to a precursor formula of the general form

$$L_{i,j} = \frac{f_{i,j}}{f_{i,j}+k} \text{ for some } k > 0 \tag{6}$$

where (6) has been described as an approximation of a mixture of two Poisson distributions; i.e. as an approximation of a 2-Poisson Model (Robertson & Walker, 1994; Baayen, 1993; Church & Gale, 1995; Rennie, 2005; Ogura, Amano, & Kondo, 2013).

Revisiting Table 2, the Box-Cox transformation for $p$ = -1 gives

$$L_{i,j} = 1 - \frac{1}{f_{i,j}+k} = \frac{f_{i,j}+k-1}{f_{i,j}+k} = \frac{f_{i,j}}{f_{i,j}+k} + \frac{k-1}{f_{i,j}+k} \tag{7}$$

which is the same as applying the inverse regression, sometimes called "reciprocal" transformation (Penn State, 2016),

$$y = \beta_0 + \beta_1 \frac{1}{x} \tag{8}$$

where

$$y = L_{i,j}$$
$$x = f_{i,j} + k$$
$$\beta_0 = 1$$
$$\beta_1 = -1$$

(8) describes an inverse regression curve that is increasing asymptotic for $\beta_1 < 0$ and decreasing asymptotic for $\beta_1 > 0$. Clearly for $k$ = 1, (6) and (7) return the same results. Since in the formal BM25 model $k$ is $K = k_1 B$, then for $K$ = 1, $k_1 = \frac{1}{B} = \frac{1}{(1-b)+b*\frac{dl}{avedl}}$ .

## Conceptual Differences between (6) and (7)

Figure 1 shows several conceptual differences between (6) and (7) for other values of $k$. As we can see, the $k$ parameter controls the rate of increase of (6) and (7), though in opposite fashions.

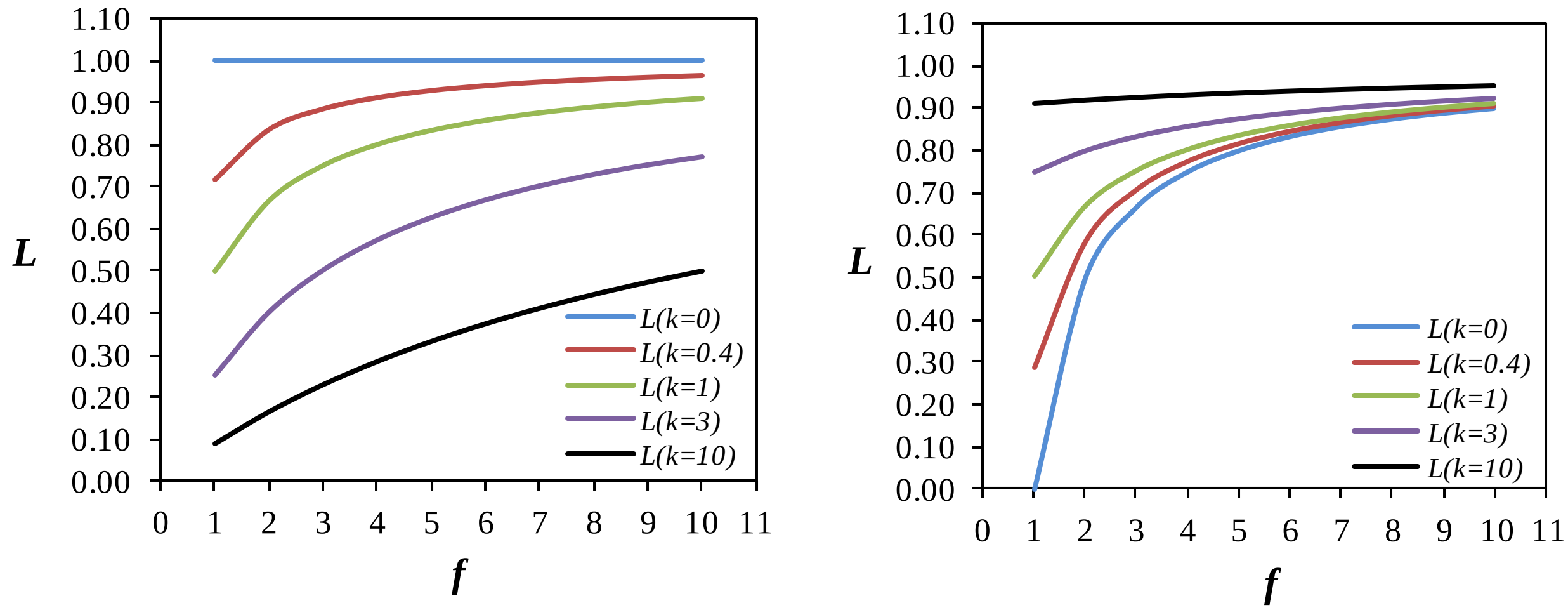

**(a) 2-Poisson approximation, (6)** **(b) Box-Cox (inverse regression), (7)**

**Figure 1. Profile curves for (a) 2-Poisson approximation and (b) Box-Cox transformation (*p* = -1).**

From Figure 1, in (a) varying $k$ from $k$ = 0 to $k$ = 10, changes the weighting behavior of (6) from binary, $L_{i,j} \begin{cases} 1 \ if \ f_{i,j} > 0 \\ 0 \ if \ f_{i,j} = 0 \end{cases}$, to almost linear. For high $k$'s increments in $f_{i,j}$ continue to contribute significantly to the local weights, whereas for low $k$'s the additional contribution of a newly observed occurrence quickly reaches a saturation point.

A different scenario is observed in (b) with (7). For high $k$'s, additional increments in $f_{i,j}$ quickly reaches a saturation regime. By contrast for low $k$'s, the newly observed occurrences contribute notably to the local weights and far more than in (a). For roughly $k$ > 0.4, (7) agrees with (6) in that the largest probabilistic evidence of eliteness occurs when a term is mentioned for the very **first time** (Robertson, 2004). More precise calculations give $k$ = 0.42 as the critical mark. Below this mark, (b) reveals that the largest weight increment and evidence of eliteness occurs when a term is mentioned for the **second time**.

Finally for $k$ = 0, (7) returns $L_{i,j} = 0$ for $f_{i,j} = 1$. This can be used to model the experimental conditions wherein one wants to reduce recall by ignoring index terms mentioned only once, effectively reducing an index term vocabulary. Alternatively, this extreme scenario can be avoided by adding a positive value to the scale of $L_{i,j}$ values or by setting $k$ to a value slightly above zero, like $k \leq 0.01$.

## A Proposed BM25 Model based on Inverse Regression

Several authors have reported reasonably good results with BM25 implementations where

- $0.5 < b < 0.8$ and $1.2 < k_1 < 2$ (BM25 classic, Robertson & Zaragoza, 2009)
- $0.3 < b < 0.6$ and $1 < k_1 < 2$ (BM25L, Lv & Zhai, 2011)

It appear from those settings that fair enough results are obtained with

$$\frac{b}{k_1} < 1 \tag{9}$$

Therefore, in this section we investigate what would be the expected profile curves in (7) by setting $k = K = k_1\left((1-b) + b * \frac{dl_j}{avedl}\right)$ for different values of (9), more specifically for $\frac{b}{k_1} \in [0.3, 0.8]$.

Replacing $k$ with $K$ leads to a BM25 weighting scheme modeled by inverse regression and that we might refer to as the Inverse Regression BM25 model or BM25IR.

$$L_{i,j} = \left(1 - \frac{1}{f_{i,j} + K}\right) = \left(\frac{f_{i,j}}{f_{i,j} + K} + \frac{K-1}{f_{i,j} + K}\right) \tag{10}$$

Figure 2 (a), (b), and (c) shows that any combination of parameters that lead to high $K$'s reduces the rate of increase of (10) for a given $f_{i,j}$. The figure corresponds to documents of (a) average, $\frac{dl_j}{avedl} = 1$, (b) short, $\frac{dl_j}{avedl} = 0.1$, and (c) long, $\frac{dl_j}{avedl} = 10$, lengths relative to the average lengths.

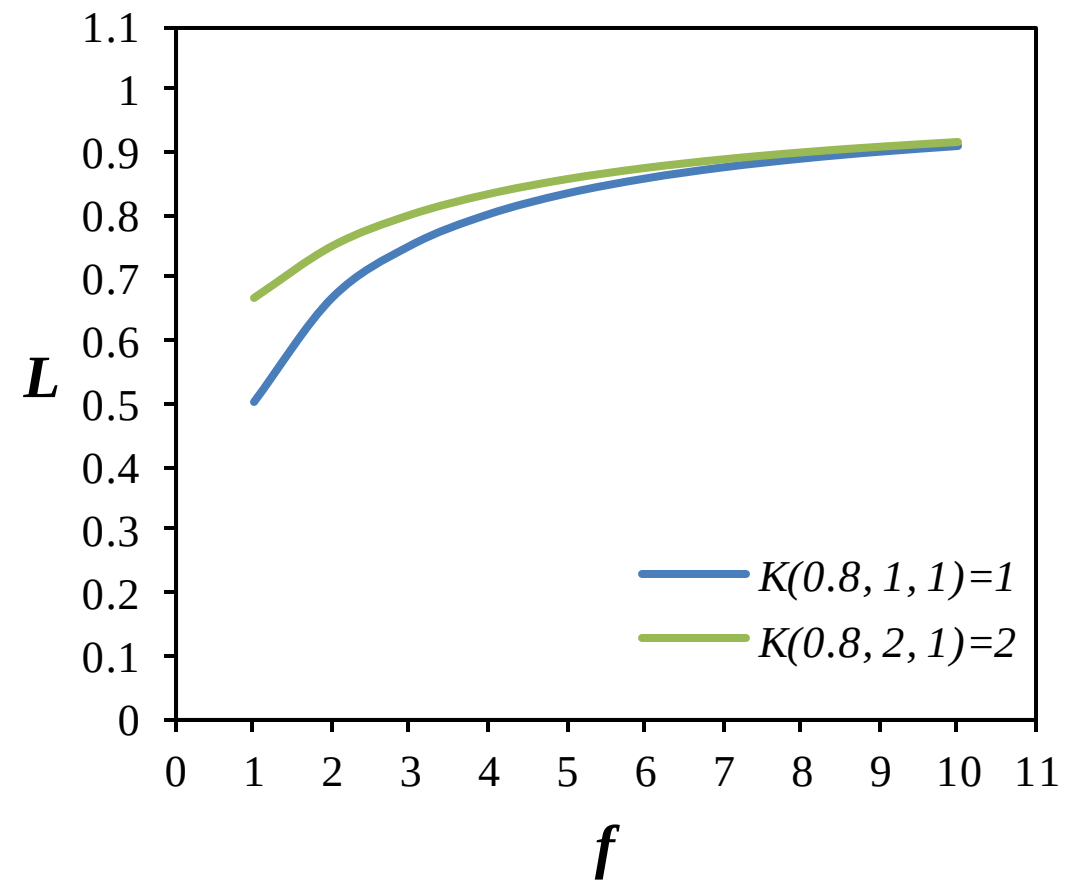

**(a)**

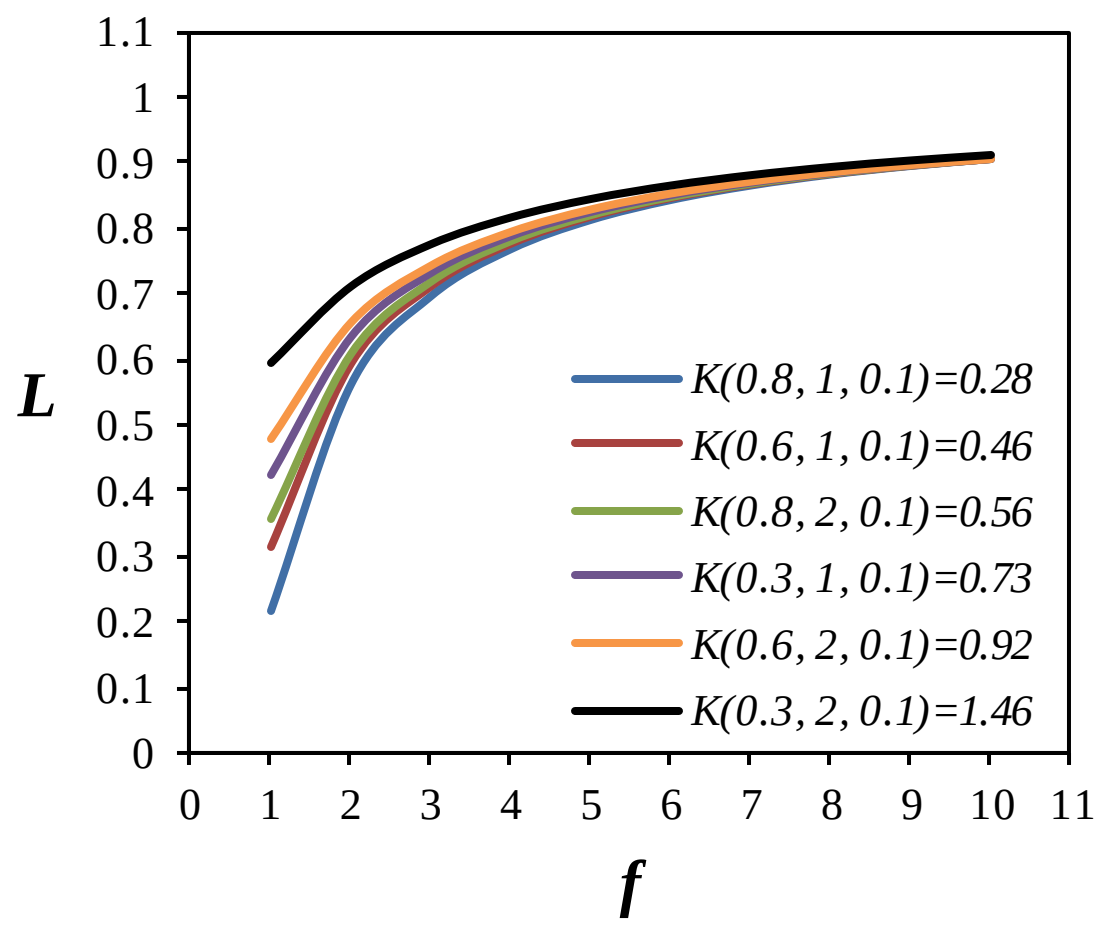

**(b)**

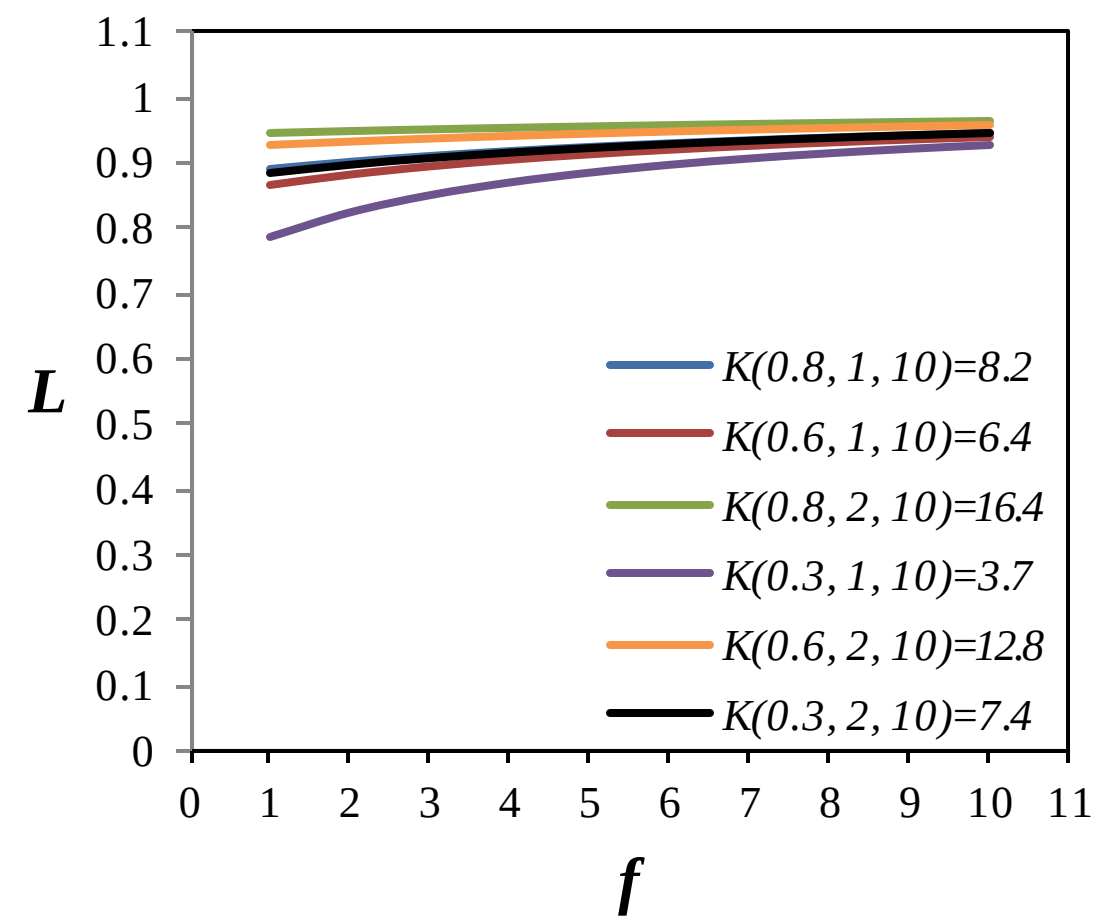

**(c)**

**Figure 2. BM25IR profile curves for different values of $K\left(b, k_1, \frac{dl_j}{avedl}\right)$.**

In (a), $K = k_1$ for any values of $b$ and $\frac{dl_j}{avedl}$ so we only show curves for $K = k_1 = 1$ and $K = k_1 = 2$. In (b) all combinations of these parameters produce acceptable results. As expected, for $K > 0.40$, the largest evidence of eliteness occurs when a term is mentioned for the first time. However, below this mark (in this case for $K = 0.28$), the largest evidence occurs when a term is mentioned for the second time.

In (c), the increment in weight is more discernible for low values of $K$, in this case for $K$ = 3.17. Thus in BM25IR, keeping $K$ low works better for the long documents considered. A comparison between (b) and (c) suggests that in BM25IR $b$ = 0.3 and $k_1$ =1 works fairly well for short and long documents, at least for the relative length considered.

Finally, for completeness and to conform BM25IR to typical BM25 schemes we may restore the scaling factor $(k_1 + 1)$, although this is not really necessary; i.e.

$$L_{i,j} = \left(1 - \frac{1}{f_{i,j} + K}\right) * (k_1 + 1) = \left(\frac{f_{i,j}}{f_{i,j} + K} + \frac{K-1}{f_{i,j} + K}\right) * (k_1 + 1) \tag{11}$$

where for $K$ = 1, BM25IR and BM25 are the same. Thus, for values of $K$ near 1, the two models should give similar, though not identical, acceptable results.

## Conclusion

Power transformations can be used as a common framework for the systematic derivation of local term weight models. We have shown through Box-Cox transformations that for $p$ = -1 and $k$ = 1, Best Match algorithms and inverse (reciprocal) regression produce equivalent results.

BM algorithms are based on the notion that the first occurrence of a query term in a document is more important than other occurrences by giving the most probabilistic evidence of eliteness. Inverse regression agrees with this assertion for $k > 0.41$. For $0 < k < 0.42$, however, inverse regression shows that the largest increment in weight and evidence of eliteness occurs when a term is mentioned for the second time.

As a result of combining BM with our inverse regression approach, we propose a new model, BM25IR, that preliminary results suggest it should work for different BM25 parametric conditions.